\documentclass[aps,preprint]{revtex4}%
\usepackage{amsfonts}
\usepackage{amsmath}
\usepackage{amssymb}
\usepackage{graphicx}
\usepackage{epstopdf}%
\providecommand{\U}[1]{\protect\rule{.1in}{.1in}}

\begin{document}
\title{Effect of an axial pre-load on the flexural vibrations of viscoelastic beams.}
\author{Elena Pierro$^{1}$}
\affiliation{$^{1}$School of Engineering, University of Basilicata, 85100 Potenza, Italy}
\pacs{PACS number}

\begin{abstract}
Polymers are ultra-versatile materials that adapt to a myriad of applications,
as they can be designed appropriately for specific needs. The realization of
new compounds, however, requires the appropriate experimental
characterizations, also from the mechanical point of view, which is typically
carried out by analyzing the vibrations of beams, but which still have some
unclear aspects, with respect to the well-known dynamics of elastic beams. To
address this shortcoming, the paper deals with the theoretical modelling of a
viscoelastic beam dynamics, and pursues the elucidation of underlying how the
flexural vibrations may be affected when an axial preload, compressive or
tensile, is applied. The analytical model presented, is able to shed light on
a peculiar behaviour, which is strongly related to the frequency dependent
damping induced by viscoelasticity. By considering as an example a real
polymer, i.e. a synthetic rubber, it is disclosed that an axial preload, in
certain conditions, may enhance or suppress the oscillatory counterpart of a
resonance peak of the beam, depending on both the frequency distribution of
the complex modulus and the length of the beam. The analytical model is
assessed by a Finite Element Model (FEM), and it turns out to be an essential
tool for understanding the dynamics of viscoelastic beams, typically exploited
to experimentally characterize polymeric materials, and which could vary
enormously simply through the application of constraints and ensued preloads.

\end{abstract}
\startpage{1}
\endpage{2}
\maketitle

\section{Introduction}

The new forthcoming technology challenges seems to be oriented towards the use
of ultra light, extremely resistant, active and super smart materials,
substantially able to face with increasingly innovative shapes
\cite{Chaudhary2021}, demands for adaptative features based on operating
conditions \cite{Terwagne2014}, more in general with self-healing properties
\cite{Wang2020} and eco-sustainable \cite{Ahmed2021}. Of particular interest,
more recently, are all those systems that aim to exploit the properties of
soft materials, taking inspiration from biological systems, which offer
countless performances, but that are also complex and therefore difficult to
replicate. For example, soft actuators \cite{Li2022}\ received great
attention, since they can improve their performance through appropriate
programming, and find applicability in the field of soft robotics
\cite{Cianchetti2018}, which seem to show excellent results in terms of
durability and reliability in the biomedical applications, and can transit
reversibly between different liquids and solids, as they swhich between
different locomotive modes \cite{Hu2018}. These latest research trends are
also part of the recently introduced concept of physical intelligence, which
in the near future will allow intelligent machines to be able to move
autonomously in various conditions of the real world \cite{Sitti2021}. As we
move towards these scenarios, already widely present in our daily life in a
vast range of applications, from the automotive to the medical field, it will
no longer be possible to use materials "fixed" in their nominal design
conditions, as they will need to be replaced by materials in constant movement
and change \cite{Rothemund2021,Martins2021}. At the moment, polymers are
between the favored materials and best suited to these circumstances, since
they can be designed to serve a specific purpose, with properly tuned physical
properties \cite{Brinson2015}, such as stiffness and damping. For this reason
they are the subject of intensive study in many engineering fields, especially
for what regards their mechanical properties, which are deeply conditioned by
viscoelasticity, as recently shown in the field of contact mechanics
\cite{Carbone2011,Carbone2012,Carbone2012bis,Carbone2013bis}. In Ref.
\cite{Pierro2020} it has been highlighted that, in particular, the
viscoelastic modulus, which exhibits a complex behaviour in the frequency
domain, is able of making the adhesion between two surfaces extremely tough or
quite weak, depending on how the imaginary part of the viscoelastic modulus is
distributed in frequency. Whether polymers are employed individually or
combined with other materials (e.g. in the case of composites), it is of
fundamental importance to suitably characterize them from a mechanical point
of view \cite{Wang2017}, for all the aforementioned applications. In fact,
numerical and theoretical predictions of the dynamics rather than the
tribological behavior of structures made of such materials, are based on their
viscoelastic response to external stresses, which depends on both frequency
and temperature, and is governed by the following stress-strain relationship
\cite{Christensen}
\begin{equation}
\sigma\left(  x,t\right)  =\int_{-\infty}^{t}G\left(  t-\tau\right)
\dot{\varepsilon}\left(  x,\tau\right)  \mathrm{d}\tau\label{stress-strain}%
\end{equation}
being $\dot{\varepsilon}(t)$ the time derivative of the strain, $\sigma(t)$ is
the stress, $G\left(  t-\tau\right)  $ is the time-dependent relaxation
function, usually characterized in the Laplace domain, through the
viscoelastic modulus $E\left(  s\right)  =sG\left(  s\right)  $. There is an
awesome quantity of research devoted to the experimental characterization of
the viscoelastic modulus $E\left(  s\right)  $, from the widespread DMA
(Dynamic mechanical analysis) technique \cite{Rasa2014}, which still presents
some problems and uncertainties, to the investigation of the dynamics of
beam-like structures \cite{Caracciolo2,Cortes2007}. In the context of this
latter experimental approach, some progress has recently been made, as in
Ref.\cite{Pierro2021}, where the vibrational response of a suspended beam
impacted with a hammer has been exploited to retrieve the complex modulus,
increasing the frequency range of interest by varying the length of the beam.
The technique is resulted reliable, accurate, and in good agreement with the
DMA. The breakthrough of the proposed technique is related to the analytical
model presented, which is able to accurately take into account, in the
vibrational response of the beam, the correct frequency trend of the
viscoelastic modulus, by varying the number of relaxation times to achieve a
good theoretical-experimental fit. However, previous theoretical studies,
focused on the dynamics of viscoelastic beam and plates
\cite{Barruetabena2012,Inman1989,Gupta2007} were lack of a specific analysis
able of linking the eigenvalues and the significant physical parameters to the
analytical response of such continuous systems, as done for example in Ref.
\cite{Adhikari2005}, for a single degree-of-freedom non-viscously damped
oscillator. \begin{figure}[ptb]
\begin{center}
\includegraphics[
height=10cm,
]{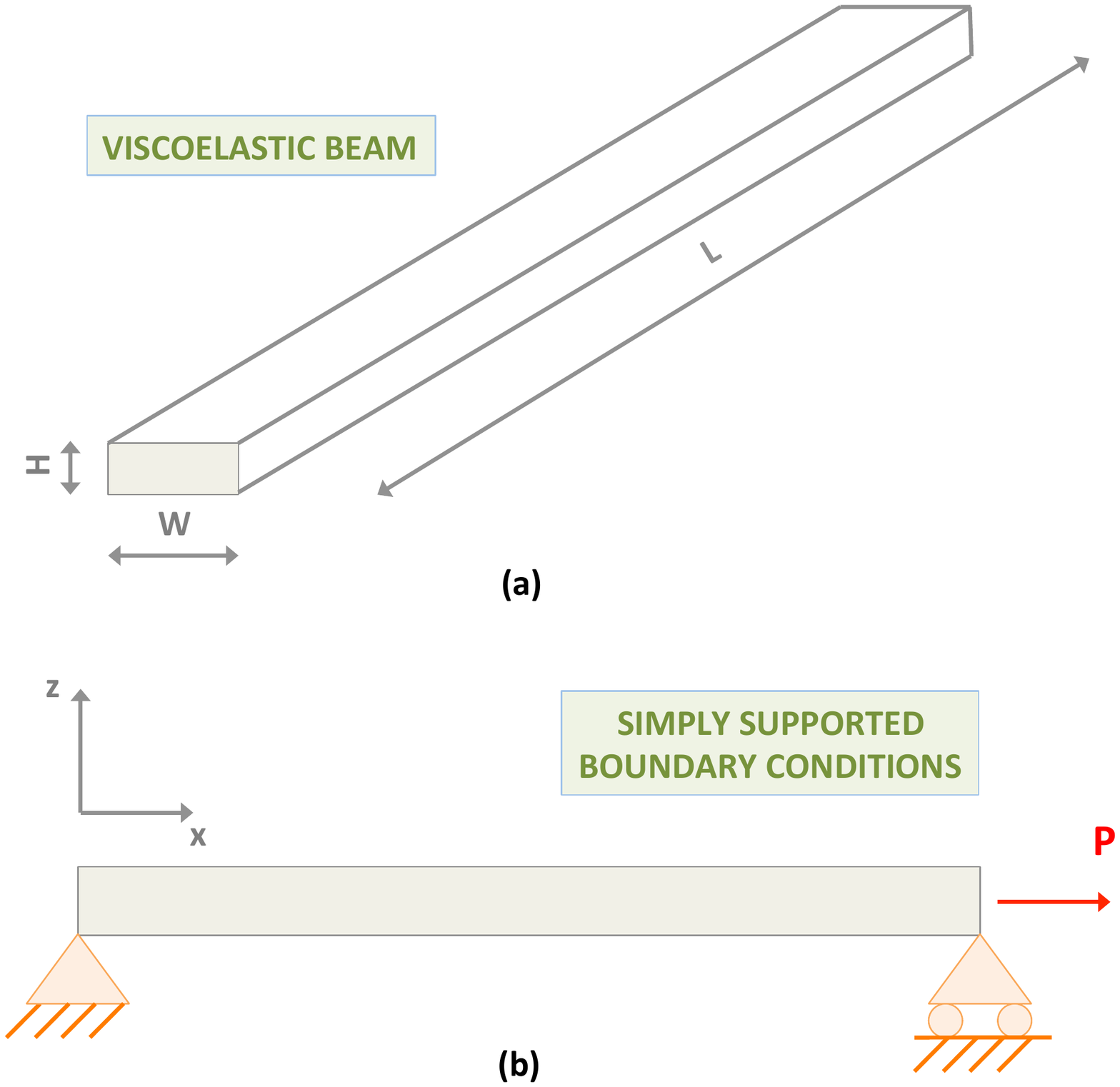}
\end{center}
\caption{The viscoelastic beam under investigation, of length $L$ and
rectangular cross section with area $A=WH$ (a), which is simply supported at
both the extremities and axially pre-loaded (b).}%
\label{Figure1}%
\end{figure}To address this shortcoming, in Ref.\cite{Pierro2020b}, some new
characteristic maps related to the nature of the eigenvalues of a viscoelastic
beam have been presented, with the aim to elucidate the influence of the
material properties and of some geometrical characteristics on the overall
beam dynamics. Interestingly, from this study it resulted that by properly
selecting the beam length, for a chosen viscoelastic material, it is possible
to suppress or enhance one resonance peak or more peaks simultaneously. This
outcome is of crucial concern for the experimental characterizations of
viscoelastic materials, as the one presented in \cite{Pierro2021}, since it
can help in accurately interpreting the resonances when shifted with different
beam lengths. Among the several possibilities to observe a further shift of
the response spectrum of the beam in the frequency domain, and therefore to
enlarge the frequency range of interest in the experimental characterization
of the viscoelastic modulus, one may i) change the surrounding temperature or
ii) apply an axial compressive/tractive pre-load to the beam. It is known
\cite{Cheli2015}, indeed, that when an elastic beam is subjected to a static
pre-load, its resonances move towards higher or lower frequencies, in case of
an applied traction or compression respectively, while no information is known
for viscoelastic beams under the same conditions. The goal of this paper is to
improve the previously presented theoretical study \cite{Pierro2020b} on the
viscoelastic beams, in order to get new insights in terms of eigenvalues, and
consequently of resonance peaks, when a tractive and a compressive pre-load
are applied. A viscoelastic material with two relaxation times is considered,
since it is always possible to divide the frequency spectrum under analysis in
several intervals, thus allowing to decrease the number of the predominant
relaxation times in such intervals \cite{Pierro2019}. The results presented
are further validated by means of a FEM analysis.

\section{Flexural vibrations of the tensioned beam}

In this section it is presented the analytical formulation to derive the
equations which governs the flexural vibrations of an axially pre-loaded
viscoelastic beam. For this scope, a homogenous beam with rectangular cross
section is considered (Figure \ref{Figure1}-a), being $L$ the length of the
beam, $W$, and $H$ the width and the thickness of the beam cross section
respectively, which are supposed to follow the slenderness condition, i.e.
$L\gg W$, $L\gg H$. Since the study presented in this paper will be centered
on the first resonances of the beam, which are not influenced by the shear
deformations, the Bernoulli theory of transversal vibrations is exploited to
describe the beam dynamics. This choice entails that the small displacements
condition along the $z$-axis, i.e. $\left\vert u\left(  x,t\right)
\right\vert \ll L$, needs to be satisfied.

The beam is supposed to be simply supported on both extremities, with an axial
pre-load $P$ applied at $x=L$ (Figure \ref{Figure1}-b). It is possible to
prove, through simple calculations \cite{Cheli2015}, that the effect of the
static axial action $P$ is introduced in the general equation of motion by
means of the term $P\partial^{2}u\left(  x,t\right)  /\partial x^{2}$. In the
case of a beam with viscoelastic properties, the equation of motion is
therefore \cite{Inman1996,Pierro2020b}%
\begin{equation}
J_{xz}\int_{-\infty}^{t}E\left(  t-\tau\right)  \frac{\partial^{4}u\left(
x,\tau\right)  }{\partial x^{4}}\mathrm{d}\tau+\mu~\frac{\partial^{2}u\left(
x,t\right)  }{\partial t^{2}}+P\frac{\partial^{2}u\left(  x,t\right)
}{\partial x^{2}}=f\left(  x,t\right)  \label{motionEq}%
\end{equation}
being $\mu=\rho A$, $\rho$ the bulk density of the material, $A=WH$ the cross
section area, $J_{xz}=(1/12)WH^{3}$ the moment of inertia, and $f\left(
x,t\right)  $ is the generic forcing term. Any other damping terms may be
added to the Eq.(\ref{motionEq}) \cite{Banks91}, such as the viscous and the
hysteretic ones, but the presented analysis is only focused on the damping
effect that comes from viscoelasticity. In order to solve Eq.(\ref{motionEq}),
the associated homogeneous problem is firstly considered
\begin{equation}
J_{xz}\int_{-\infty}^{t}E\left(  t-\tau\right)  u_{xxxx}\left(  x,\tau\right)
\mathrm{d}\tau+\mu~u_{tt}\left(  x,t\right)  -P~u_{xx}\left(  x,t\right)  =0
\label{autoproblem}%
\end{equation}
together with the boundary conditions of the simply supported beam (Figure
\ref{Figure1}-b)
\begin{align}
u\left(  0,t\right)   &  =0\label{BCautoproblem}\\
u_{xx}\left(  0,t\right)   &  =0\nonumber\\
u\left(  L,t\right)   &  =0\nonumber\\
u_{xx}\left(  L,t\right)   &  =0\nonumber
\end{align}
having posed $u_{x}\left(  x,t\right)  =\partial u\left(  x,t\right)
/\partial x$, $u_{t}\left(  x,t\right)  =\partial u\left(  x,t\right)
/\partial t$. The solution of Eq.\ref{autoproblem} can be easily found in the
Laplace domain, with initial conditions equal to zero, so that the
eigenfunctions $\phi\left(  x,s\right)  $ can be calculated solving the
equation%
\begin{equation}
\phi_{xxxx}\left(  x\right)  -P_{eq}\phi_{xx}\left(  x\right)  -\beta_{eq}%
^{4}\left(  s\right)  \phi\left(  x\right)  =0 \label{autoproblemLaplace}%
\end{equation}
with the boundary conditions%

\begin{align}
\phi\left(  0\right)   &  =0\label{BC autofunction}\\
\phi_{xx}\left(  0\right)   &  =0\nonumber\\
\phi\left(  L\right)   &  =0\nonumber\\
\phi_{xx}\left(  L\right)   &  =0\nonumber
\end{align}
having defined%
\begin{align}
\beta_{eq}^{4}\left(  s\right)   &  =-\frac{\mu~s^{2}}{J_{xz}E\left(
s\right)  }\label{beta_equivalent}\\
P_{eq}  &  =\frac{P}{J_{xz}E\left(  s\right)  }%
\end{align}
. From the characteristic equation associated to Eq.\ref{autoproblemLaplace}%

\begin{equation}
\lambda^{4}\left(  x\right)  -P_{eq}\lambda^{2}\left(  x\right)  -\beta
_{eq}^{4}\left(  s\right)  =0 \label{CharEquat}%
\end{equation}
one obtains the roots%
\begin{align}
\lambda_{a}^{2}  &  =\frac{P_{eq}-\sqrt{P_{eq}^{2}+4\beta_{eq}^{4}\left(
s\right)  }}{2}\label{lambda-a_b}\\
\lambda_{b}^{2}  &  =\frac{P_{eq}+\sqrt{P_{eq}^{2}+4\beta_{eq}^{4}\left(
s\right)  }}{2}\nonumber
\end{align}
from which%
\begin{align}
\lambda_{1,2}  &  =\pm\sqrt{\lambda_{a}^{2}}\label{lambda1-4}\\
\lambda_{3,4}  &  =\pm\sqrt{\lambda_{b}^{2}}\nonumber
\end{align}

Finally, the solution of Eq.\ref{autoproblemLaplace} can be written as%
\begin{equation}
\phi(x,s)=W_{1}\sin\left[  {\gamma}_{1}{x}\right]  +W_{2}\cos\left[  {\gamma
}_{1}{x}\right]  +W_{3}\sinh\left[  {\gamma}_{2}{x}\right]  +W_{4}\cosh\left[
{\gamma}_{2}{x}\right]  \label{eigenv}%
\end{equation}
where%
\begin{align}
{\gamma}_{1}  &  =\sqrt{-\lambda_{a}^{2}}\label{gamma12}\\
{\gamma}_{2}  &  =\sqrt{\lambda_{b}^{2}}\nonumber
\end{align}
By forcing to zero the determinant of the system matrix obtained from
Eqs.(\ref{BC autofunction}), one has the equation%
\begin{equation}
\sin\left(  \gamma_{1}L\right)  =0 \label{modes_eq}%
\end{equation}
which gives us same solutions $\gamma_{1_{n}}L=n\pi$ \cite{Inman1996} of the
elastic case. By substituting ${\gamma}_{1}^{2}=-\lambda_{a}^{2}=n\pi/L$ in
Eq.\ref{CharEquat}, the following equation can be derived%

\begin{equation}
\left(  \frac{n\pi}{L}\right)  ^{2}+P_{eq}\frac{n\pi}{L}-\beta_{eq}^{4}\left(
s\right)  =0 \label{rootEq}%
\end{equation}
from which it is possible to calculate the complex conjugate eigenvalues
$s_{n}$ corresponding to the $n_{th}$ mode, and the real poles $s_{k}$ related
to the material viscoelasticity \cite{Pierro2020b}. Furthermore, the values
$\gamma_{1n}$ allow to determine the eigenfunctions $\phi_{n}\left(  x\right)
$%
\begin{equation}
\phi_{n}\left(  x\right)  =\sin\left(  \gamma_{1n}x\right)  \label{modes}%
\end{equation}
that can be employed to get the general solution of Eq.(\ref{motionEq}),
through the decomposition \cite{Inman1989}%
\begin{equation}
u\left(  x,t\right)  =\sum_{n=1}^{+\infty}\phi_{n}\left(  x\right)
q_{n}\left(  t\right)  \label{sepvariable}%
\end{equation}
By following the same calculations shown in Ref.\cite{Pierro2020b}, and by
observing that%
\begin{align}
\phi_{n_{xx}}(x)  &  =-\gamma_{1n}^{2}\sin\left[  {\gamma}_{1}{x}\right]
=-\gamma_{1n}^{2}\phi_{n}(x)\label{eigenv_deriv}\\
\phi_{n_{xxxx}}(x)  &  =\gamma_{1n}^{4}\sin\left[  {\gamma}_{1}{x}\right]
=\gamma_{1n}^{4}\phi_{n}(x)\nonumber
\end{align}
it is straightforward to derive the projected equation of motion on the
function $\phi_{m}\left(  x\right)  $ of the basis
\begin{equation}
\mu\ddot{q}_{n}\left(  t\right)  +J_{xz}\gamma_{1n}^{4}\int_{-\infty}%
^{t}E\left(  t-\tau\right)  q_{n}\left(  \tau\right)  \mathrm{d}\tau
+\gamma_{1n}^{2}Sq_{n}\left(  t\right)  =f_{n}\left(  t\right)
\label{Eq projected time}%
\end{equation}
being $u_{m}\left(  t\right)  =\left\langle u\left(  x,t\right)  \phi
_{m}\left(  x\right)  \right\rangle =\frac{1}{L}\int_{0}^{L}u\left(
x,t\right)  \phi_{m}\left(  x\right)  \mathrm{d}x$ and $f_{n}\left(  t\right)
=\frac{1}{L}\int_{0}^{L}f\left(  x,t\right)  \phi_{n}\left(  x\right)
\mathrm{d}x$ the projected solution and the projected forcing term
respectively. By considering the Laplace Transform of
Eq.(\ref{Eq projected time}), with initial conditions equal to zero, and
forcing term equal to the Dirac Delta of constant amplitude $F_{0}$, in both
the time and the spatial domains (i.e. $f\left(  x,t\right)  =F_{0}%
\delta\left(  x-x_{f}\right)  \delta\left(  t-t_{0}\right)  $), it is possible
to obtain the system response%
\begin{equation}
U\left(  x,s\right)  =F_{0}\sum_{n=1}^{+\infty}\frac{\phi_{n}\left(  x\right)
\phi_{n}\left(  x_{f}\right)  }{\mu s^{2}+\gamma_{1n}^{2}P+J_{xz}\gamma
_{1n}^{4}E\left(  s\right)  } \label{system_response}%
\end{equation}
which clearly depends on the axial pre-load $P$.

\section{Viscoelastic model - System eigenvalues}

In order to determine the most important parameters which affect the system
dynamics, some non-dimensional quantities will be defined. For this purpose,
the general natural frequency of the transverse motions of a narrow,
homogenous beam with a bending stiffness $E_{0}J_{xz}$ and density $\rho$, is considered%

\begin{equation}
\omega_{n}=\left(  \frac{c_{n}}{L}\right)  ^{2}\sqrt{\frac{E_{0}J_{xz}}{\rho
A}} \label{general_freq}%
\end{equation}
It should be noticed that Eq.\ref{general_freq} is always valid, regardless of
the boundary conditions \cite{Thomson}, whereas the coefficient $c_{n}$
depends on the specific boundary conditions. In particular, the first natural
frequency is $\omega_{1}=\alpha^{2}\delta_{1}$, being $\delta_{1}=c_{1}%
^{2}\sqrt{E_{0}A/\left(  \rho J_{xz}\right)  }$, $\alpha=R_{g}/L$ the
dimensionless beam length, with $R_{g}=\sqrt{J_{xz}/A}$ the radius of
gyration. In the case of a rectangular beam cross section, one has
$\alpha=H/\left(  \sqrt{12}L\right)  $ and $\delta_{1}=\left(  c_{1}%
^{2}/H\right)  \sqrt{12E_{0}/\rho}$ . It is so possible to define the
non-dimensional eigenvalue $\bar{s}=s/\delta_{1}$, and in particular one has,
for the $n_{th}$ mode, $\omega_{n}^{2}=E_{0}\beta_{n}^{4}J_{xz}/\mu=r_{n}%
E_{0}$ and $\delta_{n}=c_{n}^{2}\sqrt{E_{0}A/\left(  \rho J_{xz}\right)  }$,
being $r_{n}=\left(  \beta_{n}\right)  ^{4}J_{xz}/\mu$.

Among the several constitutive models available in literature, generally
exploited to describe the stress-strain relation in Eq.\ref{stress-strain}, in
this study the generalized Maxwell model is utilized, which considers a spring
and $k$ Maxwell elements connected in parallel. The viscoelastic modulus
$E\left(  s\right)  $ in the Laplace domain, in particular, is represented by
the following discrete function%
\begin{equation}
E\left(  s\right)  =E_{0}+\sum_{k}E_{k}\frac{s\tau_{k}}{1+s\tau_{k}}
\label{ElasticModulusLaplace}%
\end{equation}
where $E_{0}$ is the elastic modulus of the material at zero-frequency,
$\tau_{k}$ and $E_{k}$ are the relaxation time and the elastic modulus
respectively of the generic spring-element in the generalized linear
viscoelastic model \cite{Christensen}. The number of relaxation times
$\tau_{k}$ typically required to well convey the complex modulus in a wide
frequency range, can be of the order of a few tens. However, it has been
recently shown that \cite{Pierro2021,Pierro2020b,Pierro2019}, in a narrow
frequency range, e.g. around a resonance peak, even just two relaxation times
are adequate for a very good representation of the modulus in that specific
range. Since the present study focuses on the analysis of some first peaks,
considered individually, and since the system is linear, the viscoelastic
modulus will be represented just through two relaxation times $\tau_{1}$ and
$\tau_{2}$. The corresponding complex function Eq.\ref{ElasticModulusLaplace},
with $k=2$, can be therefore substituted in Eq.\ref{rootEq}, and the
fourth-order characteristic equation, for each $n_{th}$ mode, can be written%
\begin{equation}
\bar{s}^{4}+\sum_{j=0}^{3}a_{j}\bar{s}^{j}=0 \label{nondim_chareq_comp_2tr}%
\end{equation}
where%
\begin{align}
a_{0}  &  =\alpha^{4}\Delta_{n}^{2}\frac{1}{\theta_{1}\theta_{2}}+\frac
{\alpha^{2}\Delta_{n}\bar{P}}{\theta_{1}\theta_{2}}\label{coeff_4orderEq}\\
a_{1}  &  =\left(  \frac{1}{\theta_{2}}+\frac{1}{\theta_{1}}+\frac{\gamma_{1}%
}{\theta_{2}}+\frac{\gamma_{2}}{\theta_{1}}\right)  \alpha^{4}\Delta_{n}%
^{2}+\frac{\alpha^{2}\Delta_{n}\bar{P}}{\theta_{2}}+\frac{\alpha^{2}\Delta
_{n}\bar{P}}{\theta_{1}}\nonumber\\
a_{2}  &  =\left(  \frac{1}{\theta_{1}\theta_{2}}+\alpha^{4}\Delta_{n}%
^{2}+\alpha^{4}\Delta_{n}^{2}\gamma_{1}+\alpha^{4}\Delta_{n}^{2}\gamma
_{2}\right)  +\alpha^{2}\Delta_{n}\bar{P}\nonumber\\
a_{3}  &  =\left(  \frac{1}{\theta_{1}}+\frac{1}{\theta_{2}}\right) \nonumber
\end{align}
having defined the non-dimensional axial pre-load $\bar{P}=P/\left(  c_{1}%
^{2}E_{0}A\right)  $, the non dimensional groups $\gamma_{1}=E_{1}/E_{0}$,
$\gamma_{2}=E_{2}/E_{0}$, $\theta_{1}=\delta_{1}\tau_{1}$, $\theta_{2}%
=\tau_{2}\delta_{1}$, and being $\Delta_{n}=\delta_{n}/\delta_{1}$. For the
quartic equation Eq.(\ref{nondim_chareq_comp_2tr}), the following discriminant
$D\left(  n\right)  $ \cite{Lazard1988}-\cite{Rees1922} can be defined%
\begin{align}
D\left(  n\right)   &  =256a_{0}^{3}-192a_{3}a_{1}a_{0}^{2}-128a_{2}^{2}%
a_{0}^{2}+144a_{2}a_{1}^{2}a_{0}-27a_{1}^{4}+144a_{3}^{2}a_{2}a_{0}^{2}%
-6a_{3}^{2}a_{1}^{2}a_{0}-80a_{3}a_{2}^{2}a_{1}a_{0}+ \label{discriminant_2tr}%
\\
&  +18a_{3}a_{2}a_{1}^{3}+16a_{2}^{4}a_{0}-4a_{2}^{3}a_{1}^{2}-27a_{3}%
^{4}a_{0}^{2}+18a_{3}^{3}a_{2}a_{1}a_{0}-4a_{3}^{3}a_{1}^{3}-4a_{3}^{2}%
a_{2}^{3}a_{0}+a_{3}^{2}a_{2}^{2}a_{1}^{2}\nonumber
\end{align}
which plays a fundamental role in the general dynamics of the beam, since it
influences the nature of the roots of Eq.(\ref{nondim_chareq_comp_2tr}). Two
of the four roots, in particular, are always real and are related to an
overdamped motion. The other two roots can be i) complex conjugate,
representing the oscillatory contribute to the $n_{th}$ mode in the beam
dynamics, or ii) both real, meaning that the $n_{th}$ mode is not oscillatory.
Finally, the acceleration of a generic beam cross-section $A\left(  x,\bar
{s}\right)  $ $=\bar{s}^{2}U\left(  x,\bar{s}\right)  $ can be written as
function of the non-dimensional parameters above defined%
\begin{equation}
A\left(  x,\bar{s}\right)  =F_{0}\sum_{n=1}^{+\infty}\frac{\bar{s}^{2}\left(
1+\theta_{1}\bar{s}\right)  \left(  1+\theta_{2}\bar{s}\right)  \phi
_{n}\left(  x\right)  \phi_{n}\left(  x_{f}\right)  }{\mu\theta_{1}\theta
_{2}\left(  \bar{s}^{4}+\sum_{j=0}^{3}a_{j}\bar{s}^{j}\right)  }
\label{SystResp_adim_2tr}%
\end{equation}
.

\section{Results}

The main results deriving from the theoretical analysis presented in this
paper, will be shown below. For the scope, the viscoelastic beam considered in
Figure \ref{Figure1} is studied when oscillating in the $xz$-plane, having a
rectangular cross section with fixed thickness $H=1~[\mathrm{cm}]$. The beam
length $L$ is considered varying by means of the parameter $\alpha=R_{g}/L$,
keeping $R_{g}=H/\sqrt{12}$ constant. Regarding the material of the beam, it
should be observed that the investigation here presented focuses the attention
on the peculiarity of polymers to be "materials in continuous change", meaning
that they see the elastic constants $E_{k}$ and the relaxation times $\tau
_{k}$ deeply changing under some operational conditions, e.g. with the
environmental temperature. In this perspective, it is not of great
significance to take fixed these constants, however a real material will be
considered as a reference, i.e. a self-adhesive synthetic rubber that has been
experimentally characterized in Ref. \cite{Desmet2015}. The elastic modulus
has been pretty well fitted by means of Eq.\ref{ElasticModulusLaplace} in
\cite{Pierro2020b}, with two relaxation times, in the frequency range
$0-10~[\mathrm{rad/s}]$, where it falls the first resonance of a beam made of
this material and with length $L=50[\mathrm{cm}]$, i.e. $\tilde{\alpha}%
=R_{g}/L=0.0058$, here considered as reference. The parameters obtained from
the fitting procedure are shown in Table 1, being $\delta_{1}=72\ast10^{3}$
for the considered boundary conditions.

\begin{center}%
\begin{tabular}
[c]{|c|c|}\hline
\multicolumn{2}{|c|}{Viscoelastic constants}\\\hline
$E_{0}=4.46\ast10^{5}~$ & $[\mathrm{Pa}]$\\\hline
$E_{1}=3.25\ast10^{6}$ & $[\mathrm{Pa}]$\\\hline
$E_{2}=1.62\ast10^{5}$ & $[\mathrm{Pa}]$\\\hline
$\tau_{1}=0.0314$ & $[\mathrm{s}]$\\\hline
$\tau_{2}=0.314$ & $[\mathrm{s}]$\\\hline
\multicolumn{2}{|c|}{$\bar{\gamma}_{1}=E_{1}/E_{0}=7.26$}\\\hline
\multicolumn{2}{|c|}{$\bar{\gamma}_{2}=E_{2}/E_{0}=0.36$}\\\hline
\multicolumn{2}{|c|}{$\bar{\theta}_{1}=\delta_{1}\tau_{1}=2267$}\\\hline
\multicolumn{2}{|c|}{$\bar{\theta}_{2}=\delta_{1}\tau_{2}=22672$}\\\hline
\end{tabular}

\bigskip{\small Table 1: Viscoelastic parameters obtained by fitting with two
relaxation times \cite{Pierro2020b} the complex modulus of the self-adhesive
rubber characterized in Ref.\cite{Desmet2015}. }
\end{center}

\begin{figure}[ptb]
\begin{center}
\includegraphics[
height=9cm,
]{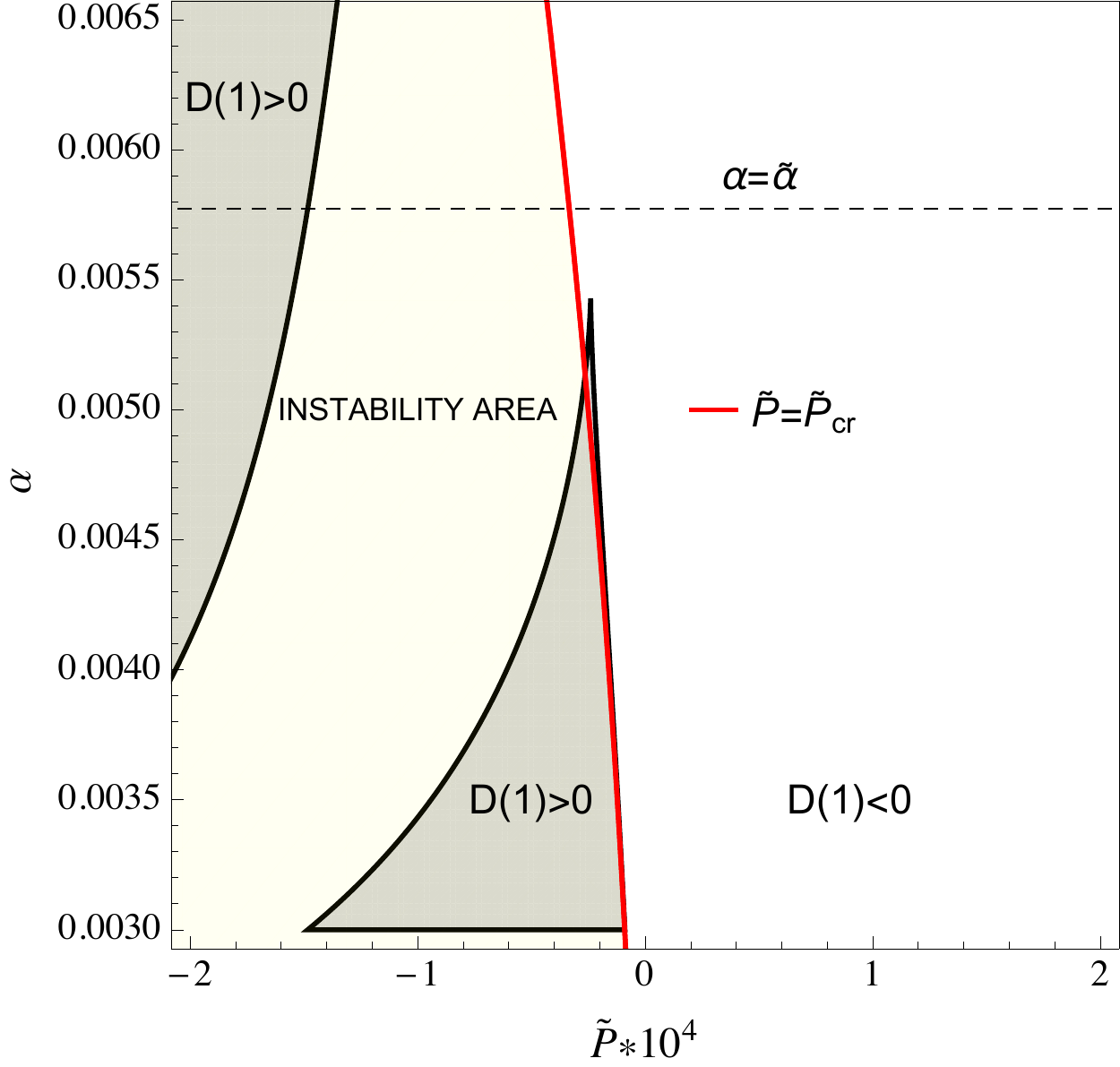}
\end{center}
\caption{The region map corresponding to the first natural frequency $n=1$,
for $\theta_{1}=$ $\bar{\theta}_{1}$, $\theta_{2}=$ $\bar{\theta}_{2}$ ,
$\gamma_{1}=$ $\bar{\gamma}_{1}$, $\gamma_{2}=$ $\bar{\gamma}_{2}$. For
$D(1)>0$ the first peak is suppressed. No tensile loads $\bar{P}>0$ determines
such condition, while for compressive load $\bar{P}<0$, the shaded areas are
almost on the left of the static Euler's critical loads calculated for every
value of $\alpha$ (red solid line), which is the area of instability.}%
\label{Figure2}%
\end{figure}

\begin{figure}[ptb]
\begin{center}
\includegraphics[
height=12cm,
]{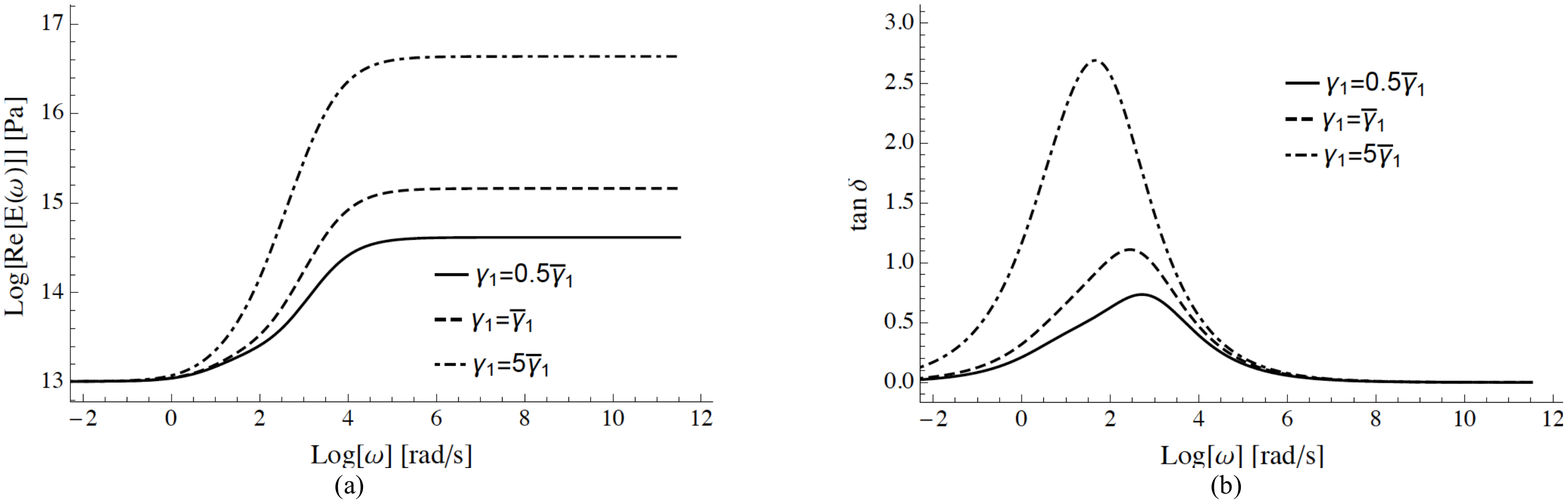}
\end{center}
\caption{The viscoelastic modulus $E\left(  \omega\right)  $, as real part
$\operatorname{Re}[E\left(  \omega\right)  ]$ (a), and the function
$\tan\delta$ (b), for $\theta_{1}=$ $\bar{\theta}_{1}$, $\theta_{2}=$
$\bar{\theta}_{2}$, $\gamma_{2}=$ $\bar{\gamma}_{2}$, and for $\gamma_{1}=$
$0.5\bar{\gamma}_{1}$ (solid lines), $\gamma_{1}=$ $\bar{\gamma}_{1}$ (dashed
lines) and $\gamma_{1}=$ $5\bar{\gamma}_{1}$ (dot dashed lines).}%
\label{Figure3}%
\end{figure}

\begin{figure}[ptb]
\begin{center}
\includegraphics[
height=12cm,
]{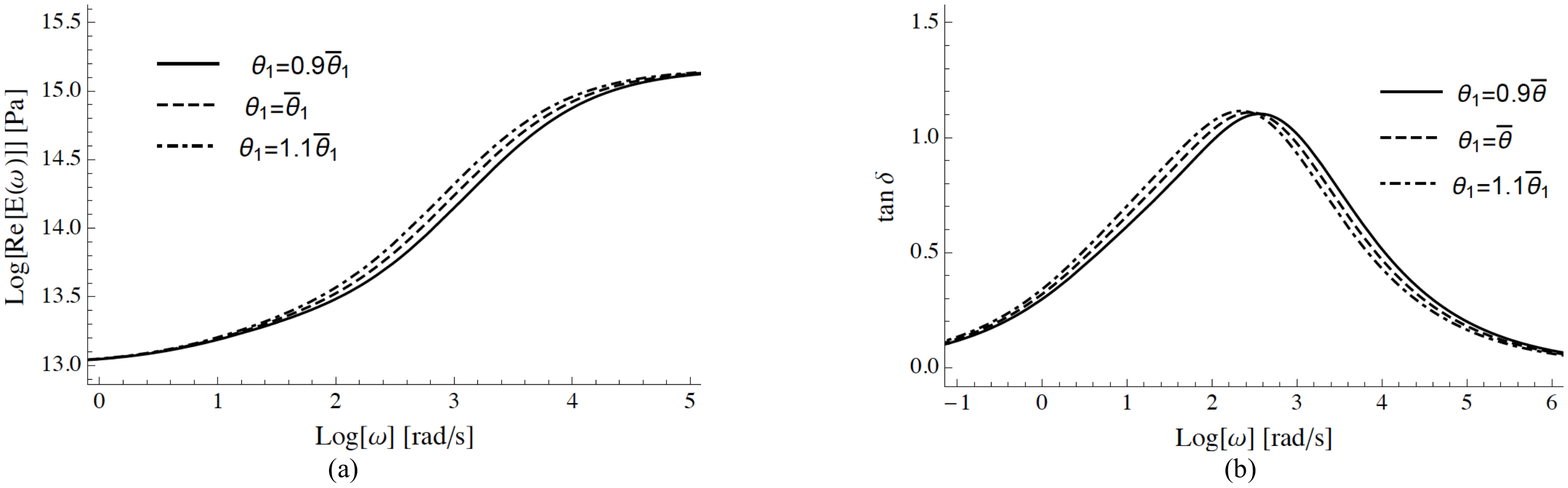}
\end{center}
\caption{The viscoelastic modulus $E\left(  \omega\right)  $, as real part
$\operatorname{Re}[E\left(  \omega\right)  ]$ (a), and the function
$\tan\delta$ (b), for $\theta_{2}=$ $\bar{\theta}_{2}$, $\gamma_{1}=$
$\bar{\gamma}_{1}$, $\gamma_{2}=$ $\bar{\gamma}_{2}$, and for $\theta_{1}=0.9$
$\bar{\theta}_{1}$ (solid lines), $\theta_{1}=$ $\bar{\theta}_{1}$ (dashed
lines) and $\theta_{1}=1.1$ $\bar{\theta}_{1}$ (dot dashed lines).}%
\label{Figure4}%
\end{figure}

In order to evaluate the effect of an axial pre-load applied to the beam, on
the first flexural mode ($n=1$), the nature of the four roots of
Eq.(\ref{nondim_chareq_comp_2tr}) is analyzed by plotting in Figure
\ref{Figure2} the discriminant $D\left(  1\right)  $
(Eq.\ref{discriminant_2tr}) as a region map, obtained by varying the parameter
values $\left(  \alpha,\bar{P}\right)  $, for $\theta_{1}=\bar{\theta}_{1}$,
$\theta_{2}=$ $\bar{\theta}_{2}$ , $\gamma_{1}=$ $\bar{\gamma}_{1}$,
$\gamma_{2}=\bar{\gamma}_{2}$. In the areas where $D(1)$ is positive, the
first peak is suppressed, but it is clear that, for the considered geometry
($\alpha=\tilde{\alpha}$) and material, there is no tensile load which
determines such condition. Even if some shaded areas with $D(1)>0$ exists for
compressive pre-loads, they are not worthy of attention, as they correspond to
loads greater than the Euler's critical load $P_{cr}=-E_{0}J_{xz}\pi^{2}%
/L^{2}$ \cite{Timoshenko1961}, which is plotted in the non-dimensional form
$\bar{P}_{cr}\left(  \alpha\right)  =P_{cr}/\left(  c_{1}^{2}E_{0}A\right)  $
in Figure \ref{Figure2} (red curve), as a function of the parameter $\alpha$,
thus delimiting the region of instability (yellow shaded area).
It is now interesting to understand if any variation of the
viscoelastic modulus, due to i) a change in the composition of the internal
material compound, or to ii) a surrounding temperature variation, with a
consequent shift of the complex modulus in the frequency domain, may somehow
affect the nature of the roots, for one or more resonance peaks. The first
condition is studied by considering, for example, the change of the constant
$E_{1}$, i.e. by varying the parameter $\gamma_{1}$, as shown in Figure
\ref{Figure3}, where the viscoelastic modulus $E\left(  \omega\right)  $ is
plotted, in terms of the real part $\operatorname{Re}[E\left(  \omega\right)
]$ (Figure \ref{Figure3}-a) and the function $\tan\delta=\operatorname{Im}%
[E\left(  \omega\right)  ]/\operatorname{Re}[E\left(  \omega\right)  ]$
(Figure \ref{Figure3}-b), for different values of $\gamma_{1}$. It is possible
to observe that by increasing $\gamma_{1}$, both the real part and the damping
contribute, represented by the function $\tan\delta$, tend to increase. The
influence of the working temperature change, which determines a frequency
shift of both the real part and the imaginary part of the complex modulus
$E\left(  \omega\right)  $, is analyzed by varying the first relaxation time
$\tau_{1}$, i.e. by changing the parameter $\bar{\theta}_{1}$. In Figure
\ref{Figure4}, in fact, one can see that an increase of $\theta_{1}$ just
determines a shift of both the real part $\operatorname{Re}[E\left(
\omega\right)  ]$ (Figure \ref{Figure4}-a) and the function $\tan\delta$
(Figure \ref{Figure4}-b), towards lower frequencies, without affecting the
amount of both the damping, i.e. the higher values of the function $\tan
\delta$, and the real part of the complex modulus.

\bigskip

\begin{figure}[ptb]
\begin{center}
\includegraphics[
height=12cm,
]{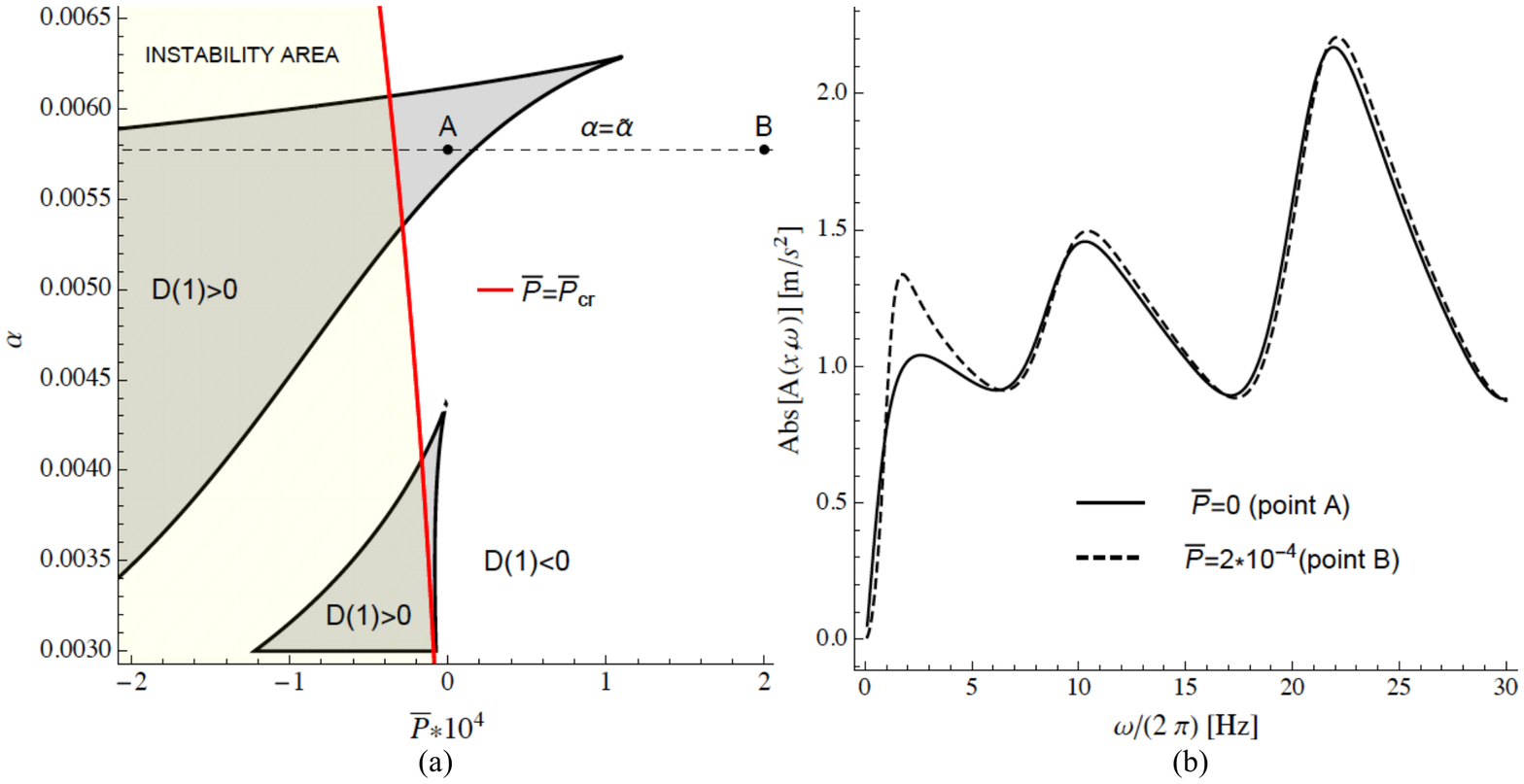}
\end{center}
\caption{The region map corresponding to the first natural frequency $n=1$,
for $\theta_{1}=$ $\bar{\theta}_{1}$, $\theta_{2}=$ $\bar{\theta}_{2}$ ,
$\gamma_{1}=5$ $\bar{\gamma}_{1}$, $\gamma_{2}=$ $\bar{\gamma}_{2}$. In this
case, the discriminant is positive $D(1)>0$ also for tractive pre-loads
$\bar{P}>0$ (e.g. in point $B$), and the suppression of the first flexural
mode can be observed when no preload is applied (e.g. in point $A$). The
static Euler's critical load is also represented (red solid line), which
delimits the instability area on the left (a); The acceleration modulus
$\left\vert A\left(  \bar{x},\omega\right)  \right\vert $ of the viscoelastic
beam is represented in frequency, in the section $x=x_{f}=\bar{x}=0.4L$, for
$\theta_{1}=$ $\bar{\theta}_{1}$, $\theta_{2}=$ $\bar{\theta}_{2}$ ,
$\gamma_{1}=5$ $\bar{\gamma}_{1}$, $\gamma_{2}=$ $\bar{\gamma}_{2}$, and for
loads $\bar{P}=0$ (point $A$) and $\bar{P}=2\ast10^{-4}$ (point $B$). This
time, only with a tensile positive pre-load (black dashed line, corresponding
to point $B$), the first resonance is clearly present (b).}%
\label{Figure5}%
\end{figure}

Focusing the attention again on the first flexural mode ($n=1$), the region
map of the discriminant $D\left(  1\right)  $ is plotted in Figure
\ref{Figure5}-a, for the same numerical values used in Figure \ref{Figure2},
except for $\gamma_{1}$, which is now considered equal to $\gamma_{1}=5$
$\bar{\gamma}_{1}$. It is clear that in this case, the shaded areas,
corresponding to the condition $D(1)>0$, hence to the first peak suppression,
regards also the positive tractive loads. This circumstance can be better
highlighted by representing the system response in two points, $A$ and $B$,
for $\alpha=\tilde{\alpha}$, without pre-tension $\bar{P}=0$ (point $A$) and
for a tractive pre-load $\bar{P}=2\ast10^{-4}$ (point $B$). In Figure
\ref{Figure5}-b, the acceleration modulus $\left\vert A\left(  \bar{x}%
,\omega\right)  \right\vert $ (Eq.\ref{SystResp_adim_2tr}), evaluated at the
beam section $x=x_{f}=\bar{x}=0.4L$, is shown for the two points of Figure
\ref{Figure5}-a, $A$ and $B$. It is quite clear that the beam presents a first
mode suppression, when no axial load is applied (point $A$, black solid line).
However, when the beam is pre-loaded through a tensile load $\bar{P}%
=2\ast10^{-4}$ (point $B$, black dashed line), which corresponds to a force
$P\simeq1[\mathrm{N}]$, the first mode becomes again oscillatory, and a peak
close to $10~[\mathrm{rad/s}]$ is well visible.\begin{figure}[ptb]
\begin{center}
\includegraphics[
height=6cm,
]{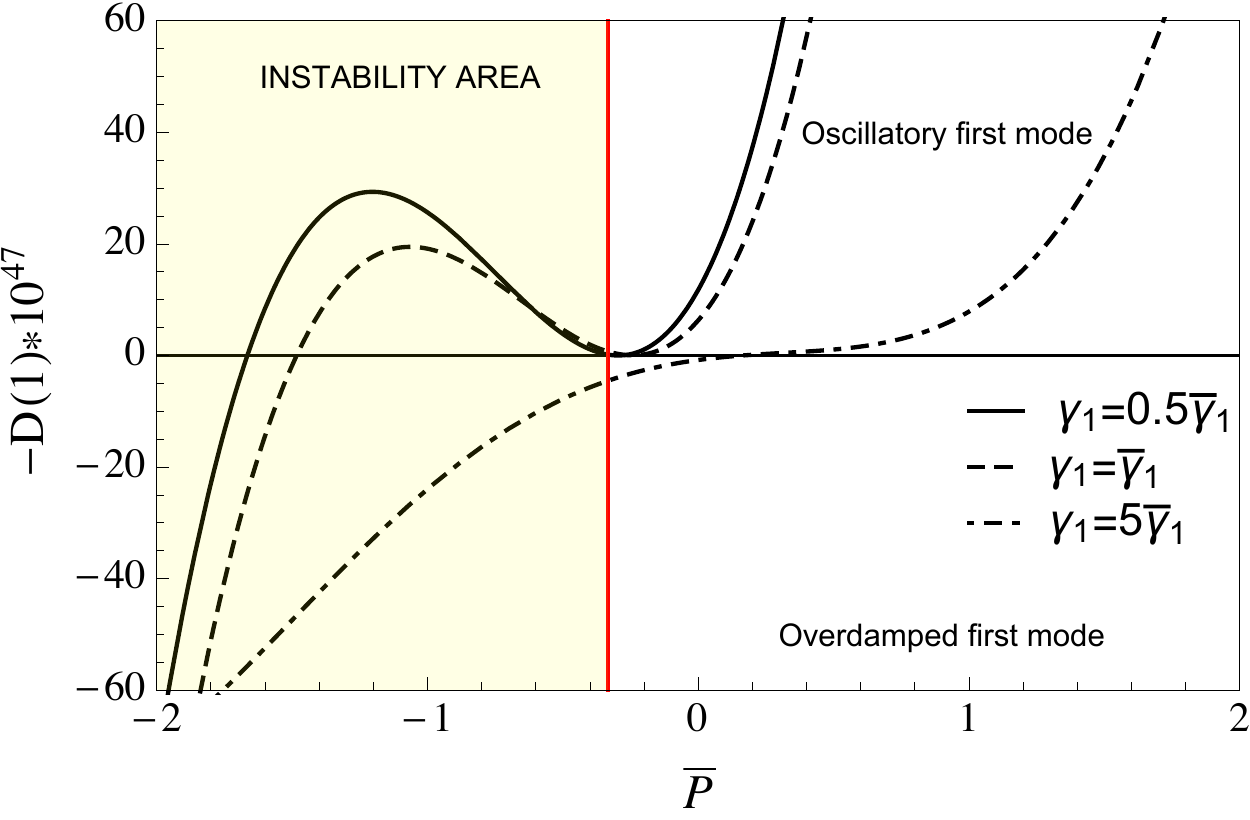}
\end{center}
\caption{The discriminant $D\left(  1\right)  $, for the first natural
frequency $n=1$, as a function of the non dimensional pre-load $\bar{P}$, for
$\alpha=\tilde{\alpha}$, $\theta_{1}=$ $\bar{\theta}_{1}$, $\theta_{2}=$
$\bar{\theta}_{2}$ , $\gamma_{2}=$ $\bar{\gamma}_{2}$, and for different
values of $\gamma_{1}$, i.e. $\gamma_{1}=$ $0.5\bar{\gamma}_{1}$ (solid line),
$\gamma_{1}=$ $\bar{\gamma}_{1}$ (dashed line) and $\gamma_{1}=$ $5\bar
{\gamma}_{1}$ (dot dashed line). The red line corresponds to the Euler's
critical load, in this case equal to $\bar{P}_{cr}\simeq-0.33$.}%
\label{Figure6}%
\end{figure}\begin{figure}[ptb]
\begin{center}
\includegraphics[
height=6cm,
]{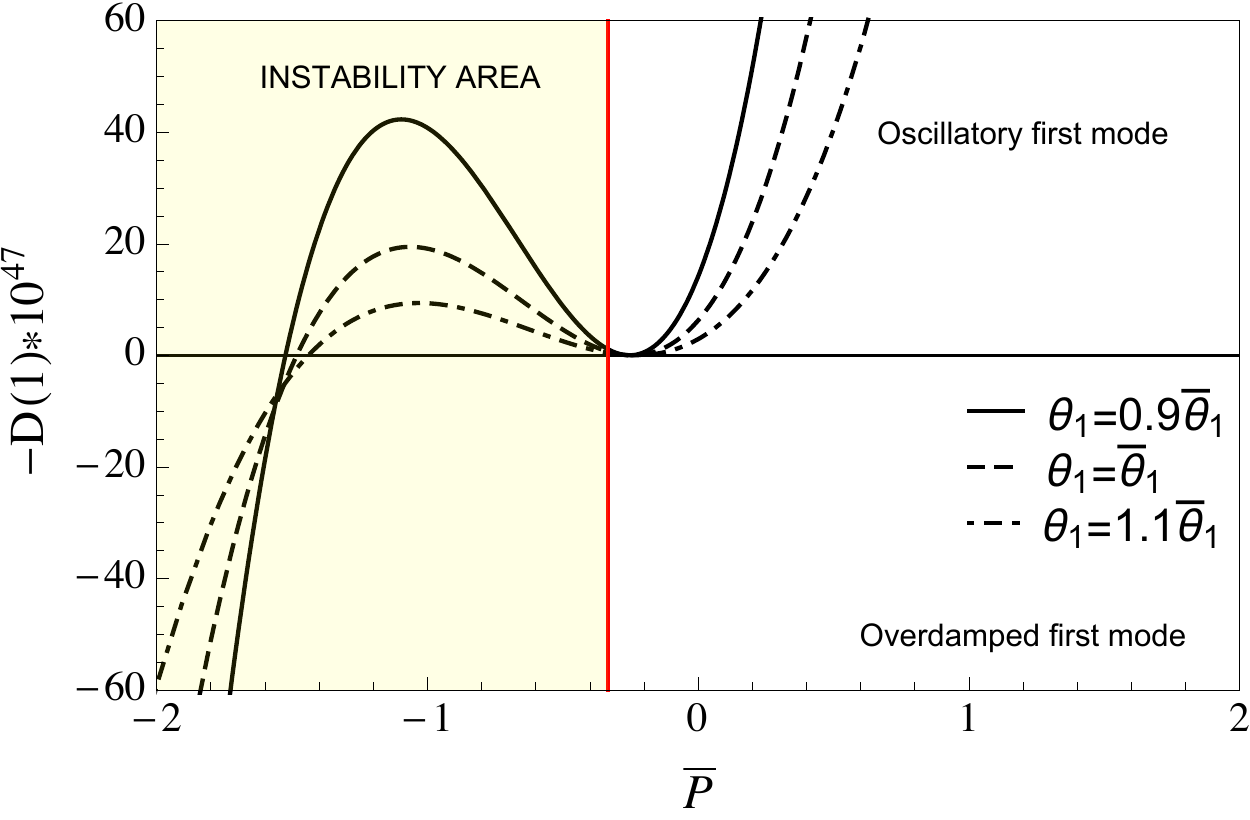}
\end{center}
\caption{The discriminant $D\left(  1\right)  $, for the first natural
frequency $n=1$, as a function of the non dimensional pre-load $\bar{P}$, for
for $\alpha=\tilde{\alpha}$, $\theta_{2}=$ $\bar{\theta}_{2}$, $\gamma_{1}=$
$\bar{\gamma}_{1}$, $\gamma_{2}=$ $\bar{\gamma}_{2}$, and for $\theta_{1}=0.9$
$\bar{\theta}_{1}$ (solid line), $\theta_{1}=$ $\bar{\theta}_{1}$ (dashed
line) and $\theta_{1}=1.1$ $\bar{\theta}_{1}$ (dot dashed line). The red line
corresponds to the Euler's critical load, in this case equal to $\bar{P}%
_{cr}\simeq-0.33$.}%
\label{Figure7}%
\end{figure}To better understand the influence of the parameters $\gamma_{1}$
and $\theta_{1}$ on the nature of the system roots, and in particular the
behaviour of the viscoelastic beam at its first natural frequency, the
discriminant $D\left(  1\right)  $ is shown as a function of the pre-tension
$\bar{P}$, at the fixed beam length $\alpha=\tilde{\alpha}$, for different
values of $\gamma_{1}$ (Figure \ref{Figure6}) and $\theta_{1}$ (Figure
\ref{Figure7}). For the particular case considered, in terms of geometrical
and material properties, and hence beam length, it is quite evident in Figure
\ref{Figure6}, again, that an increase of $\gamma_{1}$, i.e. for $\gamma
_{1}=5$ $\bar{\gamma}_{1}$, the first resonance is suppressed also in absence
of pre-load, and that tensile pre-loads could rehabilitate the oscillatory
motion of the beam at its first natural frequency. On the contrary, the motion
is always oscillatory for any variation of $\theta_{1}$, as shown in Figure
\ref{Figure7}, except for slight compressive loads, up to the Euler's critical
load $\bar{P}_{cr}\left(  \alpha=\tilde{\alpha}\right)  \simeq-0.33$%
.\begin{figure}[ptb]
\begin{center}
\includegraphics[
height=12cm,
]{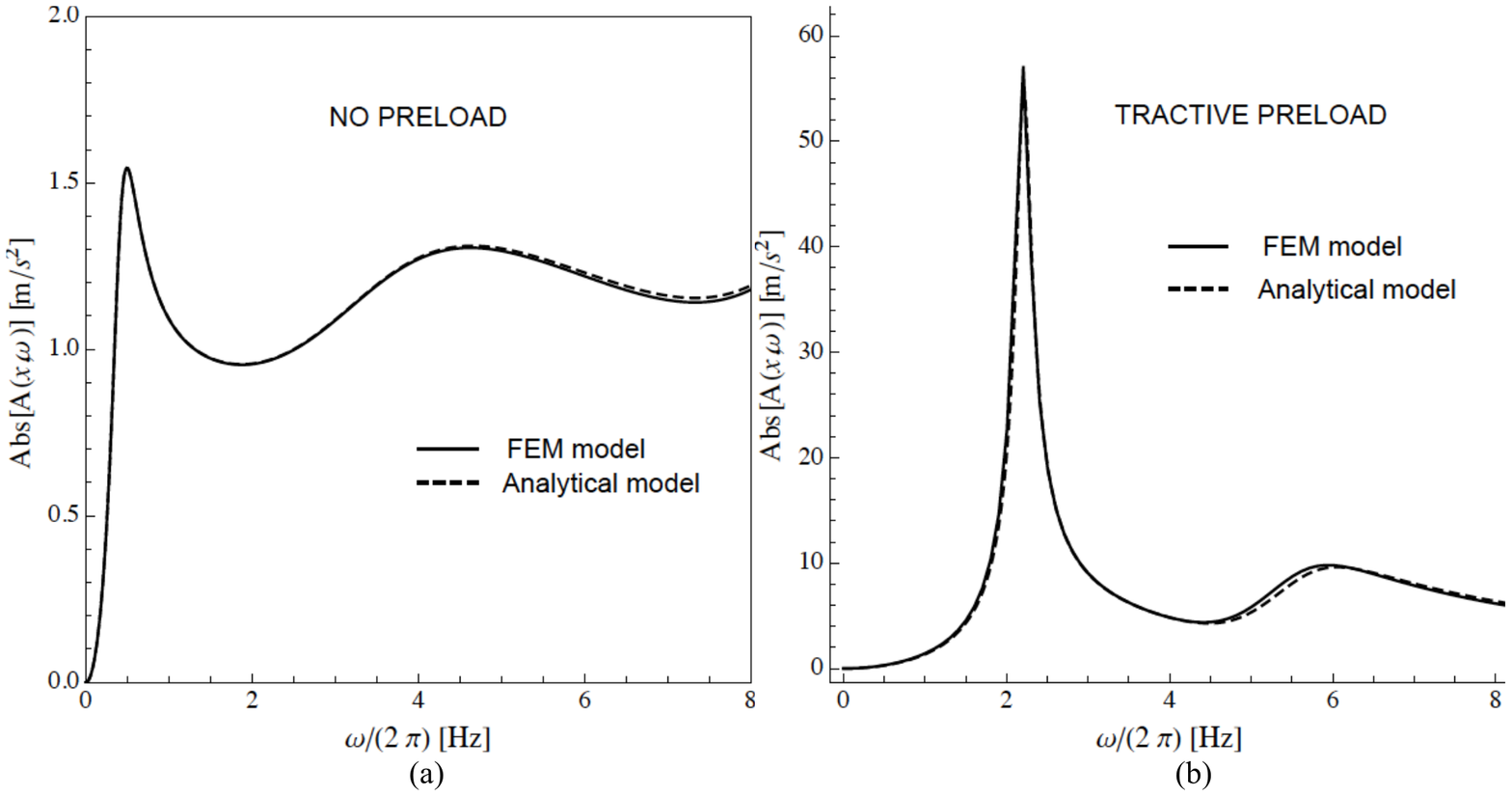}
\end{center}
\caption{The acceleration modulus $\left\vert A\left(  \bar{x},\omega\right)
\right\vert $ of the viscoelastic beam in the section $x=x_{f}=\bar{x}=0.4L$,
for $\theta_{1}=$ $\bar{\theta}_{1}$, $\theta_{2}=$ $\bar{\theta}_{2}$ ,
$\gamma_{1}=\bar{\gamma}_{1}$, $\gamma_{2}=$ $\bar{\gamma}_{2}$, in absence of
pre-load $\bar{P}=0$ (a) and for a tractive pre-load $\bar{P}=2\ast10^{-4}$
(b). In both the cases, a good agreement has been achieved, between the FEM
analysis (solid lines)\ and the theoretical model (dashed lines).}%
\label{Figure8}%
\end{figure}\begin{figure}[ptb]
\begin{center}
\includegraphics[
height=12cm,
]{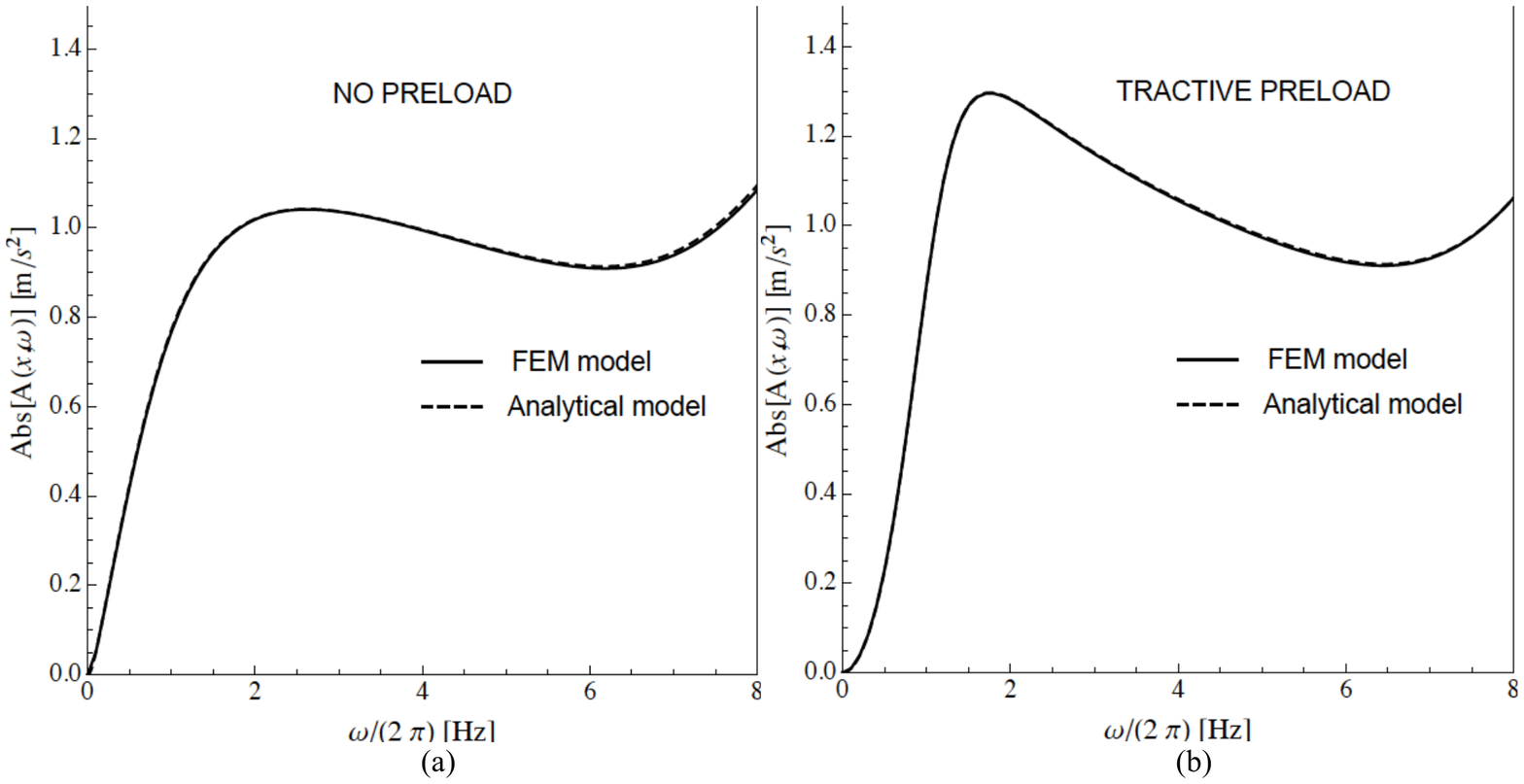}
\end{center}
\caption{The acceleration modulus $\left\vert A\left(  \bar{x},\omega\right)
\right\vert $ of the viscoelastic beam in the section $x=x_{f}=\bar{x}=0.4L$,
for $\theta_{1}=$ $\bar{\theta}_{1}$, $\theta_{2}=$ $\bar{\theta}_{2}$ ,
$\gamma_{2}=$ $\bar{\gamma}_{2}$, this time with $\gamma_{1}=5\bar{\gamma}%
_{1}$, in absence of pre-load $\bar{P}=0$ (a) and for a tractive pre-load
$\bar{P}=2\ast10^{-4}$ (b). Also in this case, it is possible to ascertain the
good agreement between the FEM analysis (solid lines) and the theoretical
model (dashed lines).}%
\label{Figure9}%
\end{figure}

\subsection{FEM simulation and final remarks}

The beam under investigation, of length $L=50[\mathrm{cm}]$, i.e.
$\tilde{\alpha}=0.0058$, and material properties reported in Table 1, has been
modelled in Abaqus \cite{Abaqus2018} by means of 6400 solid linear hexahedron
elements type (C3D8). The boundary conditions have been applied at the two
extremities, at the middle plane of the beam, to simulate the simply supported
BC. A constant force in the frequency domain, with unit amplitude, has been
applied at the beam section $x_{f}=0.4L$, where the beam acceleration has been
calculated ($x=x_{f}=0.4L$), through the steady-state dynamics module. In
Figure \ref{Figure8}, the acceleration modulus $\left\vert A\left(  \bar
{x},\omega\right)  \right\vert $ is plotted near the first natural frequency,
when no static pre-load is applied Figure (\ref{Figure8}-a), and in presence
of a tractive pre-load $\bar{P}=2\ast10^{-4}$ (Figure \ref{Figure8}-b), for
both the models, numerical (solid lines) and analytical (dashed lines). The
agreement between the two models is well established, and the considerable
increase of the acceleration amplitude due to the application of a tensile
load (Figure (\ref{Figure8}-b) is quite congruent with the region map shown in
Figure \ref{Figure2}, which foresees a low peak in the absence of pre-load,
because we are close to the area with a positive discriminant $D\left(
1\right)  >0$. In the case of applied pre-load, on the other hand, we are very
far from the area of the oscillatory motion suppression, and the peak is
particularly enhanced. Furthermore, in Figure \ref{Figure9} the acceleration
modulus $\left\vert A\left(  \bar{x},\omega\right)  \right\vert $ is shown for
the same beam and the same material, except for the parameter $\gamma_{1}$,
which is now taken $\gamma_{1}=5$ $\bar{\gamma}_{1}$. For both the cases, i.e.
in absence of pre-load (Figure \ref{Figure9}-a) and in presence of a static
tension $\bar{P}=2\ast10^{-4}$ (Figure \ref{Figure9}-b), the results coming
from the theoretical model presented in this paper, follow pretty well the
curves obtained by the FEM\ analysis, and the reduced amplitude of the first
peak is again in agreement with what has been argued about the Figure
\ref{Figure5}.

In conclusion, through the proposed analytical model, which now takes into
account the presence of a static pre-load acting on the viscoelastic beam, it
is possible to fully evaluate the dynamic response of this kind of system,
which strongly differs from the case of a perfectly elastic beam, because of
viscoelasticity. The enhancement or the suppression of a resonance peak, which
occurs only by slightly varying an axial pre-load and that, in particular
conditions, can also be involuntary and due to the effective application of
the constraints in the experimental activities, is strategic in the context of
the characterization of such materials. In in the most popular classical
techniques, such as the DMA, the accurate positioning of the constraints on
the beam can be decisive in order to retrieve the correct viscoelastic
constants. Furthermore, in the more recently proposed experimental method
\cite{Pierro2021}, where the resonance peaks are moved in the frequency
spectrum by changing the beam length, with the aim to increase the range of
interest under investigation, the controlled application of an axial pre-load
may be strategic to further increase the width of the frequency range.
Finally, the study here presented discloses aspects on polymers not
highlighted so far, which further position them among the most versatile and
tunable materials, crucial for all current and future applications.

\section{Conclusions}

In this work an analytical model has been proposed which is able to accurately
describe the transversal dynamics of viscoelastic beams, also taking into
account the effect of axial pre-loads. The main purpose is to evaluate how
these pre-loads determine a variation of the nature of the system's
eigenvalues, and therefore on the type of vibrational motion of the beam at a
certain resonance frequency. Because of the viscoelasticity, and the related
damping distribution on frequency, the behaviour of the beam is not as simple
and predictable as in the case of perfectly elastic beams. By applying a
tensile or a compressive axial pre-load, one may observe the enhancement or
the mitigation of a resonance peak, but this circumstance is incidental to a
pivotal geometrical parameter, i.e. the beam length. Same observations have
been made through a FEM analysis, which has provided results perfectly in
agreement with those obtained from the analytical model. This theoretical
model has made it possible to get new insights on how the mechanical
characteristics of polymers can completely change the dynamic behavior of a
beam. On one hand these findings are essential for all experimental
applications that make use of beams to characterize the complex viscoelastic
module, on the other they further point out the versatility of polymers, and
how they increasingly reflect the perfect peculiarities that are required by
the materials of the future.

\end{document}